\documentclass[%
 reprint,
superscriptaddress,
 amsmath,amssymb,
 aps,
]{revtex4-1}

\usepackage{graphicx}  
\usepackage{dcolumn}   
\usepackage{bm}        
\usepackage{amssymb}   
\usepackage{amsmath}
\usepackage{latexsym}
\usepackage{epsfig}
\usepackage{amsbsy}
\usepackage{array}
\usepackage{setspace}
\usepackage{bm}
\usepackage{braket}
\usepackage{multirow}
\usepackage{float}
\usepackage{subfigure}
\usepackage{subfig}
\usepackage{caption}
\usepackage{color}
\begin{document}
\preprint{APS/123-QED}
\title{Ferromagnetism and d+id superconductivity in 1/2 doped correlated systems on triangular lattice}

\author{Bing Ye, Andrej Mesaros and Ying Ran}
\affiliation{Boston College}

\date{\today}

\begin{abstract}
We investigate the quantum phase diagram of t-J model on triangular lattice at 1/2 doping with various lattice sizes by using a combination of density matrix renormalization group (DMRG), variational Monte Carlo and quantum field theories. To sharply distinguish different phases, we calculated the symmetry quantum numbers of the ground state wave functions, and the results are further confirmed by studying correlation functions. Our results show there is a first order phase transition from ferromagnetism to d+id superconductivity, with the transition taking place at $J/t=0.4\pm0.2$.
\end{abstract}

\pacs{Valid PACS appear here}
\maketitle

{\section{\label{sec:level1} Introduction}} 
The triangular lattice is the building block of many transition metal oxides \cite{1, PhysRevB.91.165139, PhysRevB.91.155135, PhysRevLett.78.1323} and organic salts \cite{2-1,2-2,Yamashita2008, doi:10.1143/JPSJ.67.3691}, in which it adds geometric frustration to the interacting electrons. Correlated electronic systems on the triangular lattice have attracted considerable attention\cite{3-1,3-2}, and interesting quantum phases have been revealed in a number of materials including unconventional superconductivity \cite{4, PhysRevB.68.104508} and quantum spin liquids \cite{5, Yamashita2009}. 

It remains challenging to theoretically understand the quantum phase diagram of such correlated electronic systems, in particular in the presence of doping. It is nevertheless known that the interplay between different competing orders, e.g., superconductivity and magnetism, could play a crucial role \cite{PhysRevB.50.647}. 

In this work we consider the t-J model on the triangular lattice, which has been an especially useful model to describe many transition metal oxides:
\begin{equation}
\label{hamiltonian}
\begin{aligned}
H_{tJ} & =P_G\sum_{<i,j>,\alpha}-t(c_{i\alpha}^\dagger c_{j\alpha}+h.c.)P_G \\ & +P_G\sum_{i,j} J (\textbf{S}_i \cdot \textbf{S}_j-\frac{1}{4} n_i \cdot n_j)P_G. \\
\end{aligned}
\end{equation}
Here $P_G$ is the Gutzwiller projection that projects out the double occupancies in t-J model, $c_{i\alpha}$ labels the annihilation operator, $\alpha$ denotes the spin index, $\textbf{S}_i$ and $n_i$ label the spin and density operators on site $i$ respectively. We will consider particular commensurate fillings, corresponding to $3/2$ ($1/2$) electrons per site for the positive (negative) nearest neighbor hopping amplitude $t$. Since these two cases are related by a particle-hole transformation and thus can be treated simultaneously, below we focus on the positive $t$ case.

\begin{figure}[t]
\centering
\includegraphics[width=3.4in]{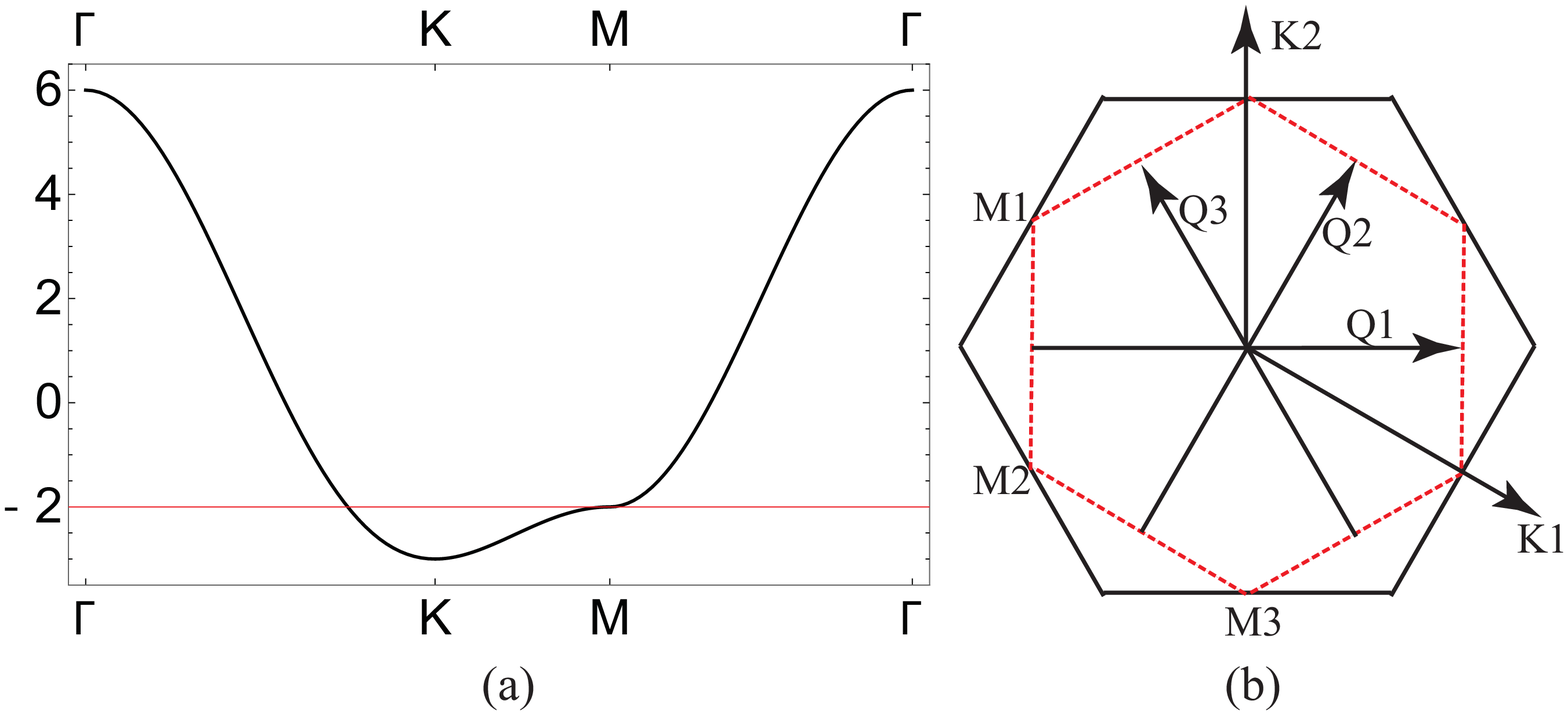}
\caption{ (a) Band structure of the nearest neighbor tight-binding model after particle-hole transformation ($t\to-t<0$). The red line is the Fermi level at 1/2 (hole) doping. (b) The nested Fermi surface (red dotted hexagon) and the Brillouin Zone (black hexagon) of this model.\label{fs}}
\end{figure} 

At this particular filling, the Fermi surface of the non-interacting nearest neighbor tight-binding model has two obvious features: as shown in Fig.\ref{fs}, the hexagon shaped Fermi surface is nested by three nesting wavevectors $Q_{1,2,3}$, and has three van Hove singularities located at  M-points $M_{1,2,3}$. 

Such a Fermi surface is clearly unstable even in the presence of weak interactions. A conventional mean-field analysis leads to spin-density-wave(SDW) orders at the nesting wavevector \cite{PhysRevLett.114.216402}. Among different kinds of SDW orders, a particularly interesting pattern is the so-called chiral SDW (c-SDW) which features quantized anomalous Hall effect \cite{PhysRevLett.114.216402, 6}. In addition, a recent renormalization group analysis shows that, at least for weak interactions, the ground state of the system should be a chiral d+id superconductor (SC) due to the scattering processes involving the van Hove singularities \cite{7}. 

However, the t-J model has no weak coupling limit, so it is unclear whether the weak coupling results apply, although they highlight the competition between superconductivity and magnetism in the 1/2 doped triangular lattice system. In addition, in the strongly coupled, small $J/t$ regime, (adiabatically connected to the large $U/t$ regime of the Hubbard model), it has been argued that ferromagnetism is an important competing phase at least when the doping is small \cite{8}.  This motivates us to carefully study the quantum phase diagram of the t-J model in this system.

In order to quantitatively investigate the quantum phase diagram of such a strongly correlated system, we use intensive numerical simulations which can provide the ground states without bias. However, reliably distinguishing competing quantum phases in such simulations has been a long-standing theoretical challenge. This is mainly due to the following conflict. On the one hand, numerical simulations become prohibitively demanding as system size grows. On the other hand, quantum phases are generally defined by their long-range physics, which requires measuring long-range correlation functions. But does one always need long-range physics to distinguish candidate quantum phases? The answer is no, and we take advantage of this fact.

As a trivial example, in order to distinguish a ferromagnetic phase and the spin-singlet superconductor phase, instead of measuring long-range correlators, one could simply look at the ground state spin quantum numbers even on rather small samples. The ferromagnetic phase should feature a large spin quantum number while the spin-singlet superconductor wave function should be in the spin-singlet sector. In more complicated examples, candidate quantum phases may have distinct {\em lattice} quantum numbers, which are generally nontrivial to compute yet are accessible numerically.

When two quantum phases are found to host distinct quantum numbers (lattice, spin, or other quantum numbers) on a sequence of finite size samples up to the thermodynamic limit, they are distinct in their short-range physics. We distinguish quantum phases in numerical simulations by comparing quantum numbers in a sequence of smaller system sizes, without having to perform the challenging finite size scaling of correlators in larger system sizes.

In this work, we study the phase diagram of the model systems using a combination of analytical construction of symmetric wave functions, the density matrix renormalization group(DMRG) \cite{RevModPhys.77.259,9-1,9-2, doi:10.1080/00018730600766432} and the variational Monte Carlo numerical simulations \cite{10-1,10-2, PhysRevB.16.3081,PhysRevLett.79.1173}. DMRG has been shown to be a nearly unbiased numerical simulation method and has been successfully applied to strongly correlated electronic systems \cite{11,12}. The basic strategy of our method is first analytically studying the characteristic symmetry quantum numbers of candidate quantum phases, and then comparing them with numerical ground state wave functions obtained from DMRG. This allows us to distinguish the candidate phases reliably even on limited system sizes. The phase diagram is then further confirmed based on correlation function measurements and a complementary variational Monte Carlo study. Previously, this method has been successfully applied to quarter doped correlated electronic systems on the honeycomb lattice \cite{11}.

\begin{figure}[t]
\centering
\includegraphics[width=3.4in]{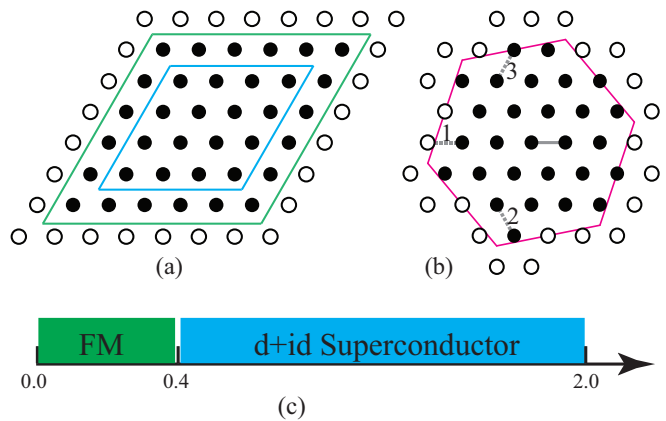}
\caption{(a) Samples with rhombus shape: 36-sites sample (bounded by green rhombus), and 16-sites sample (bounded by blue). (b) Purple hexagon bounds the 28-sites sample. Periodic boundary conditions are applied for all three samples in (a) and (b). The correlation function $\left< \hat{B}_{ij}^\dagger \hat{B}_{kl} \right>$ is chosen with bond $ij$ to be the gray solid bond, and bonds $kl$ to be the three gray dashed bonds, relabeled by index $\alpha=1,2,3$. (c) Quantum phase diagram of t-J model on triangular lattice at 1/2 doping. The ferromagnetic phase occurs at $J/t<0.4\pm0.2$, followed with increasing $J$ by a first order phase transition to a d+id superconducting phase which occurs in the regime of $0.4\pm0.2<J/t<2$.\label{samples_and_pd}}
\end{figure}

We perform DMRG simulations on 16-sites, 28-sites, and 36-sites samples, which are shown in Fig.\ref{samples_and_pd}. 


Our calculation reveals that for $J/t<0.4\pm0.2$, the system develops ferromagnetism, while with increasing J a first order phase transition into a d+id superconductor occurs. The main results are summarized in Fig.\ref{samples_and_pd}(c).


This paper is organized as follows. In Sec. \ref{sec:level1}, we construct the wave functions of c-SDW and d+id SC, and calculate the relevant symmetry quantum numbers for various system sizes, analytically and using variational Monte Carlo simulations. In Sec.\ref{sec:level2} we compare these results with DMRG analysis, including spin-spin and pair-pair correlations, to justify the phase diagram in Fig.\ref{samples_and_pd}(c).

\vfill
\section{ Wave functions of c-SDW and d+id SC }\label{sec:level1}
\begin{figure}[t]
\centering
\includegraphics[width=3.4in]{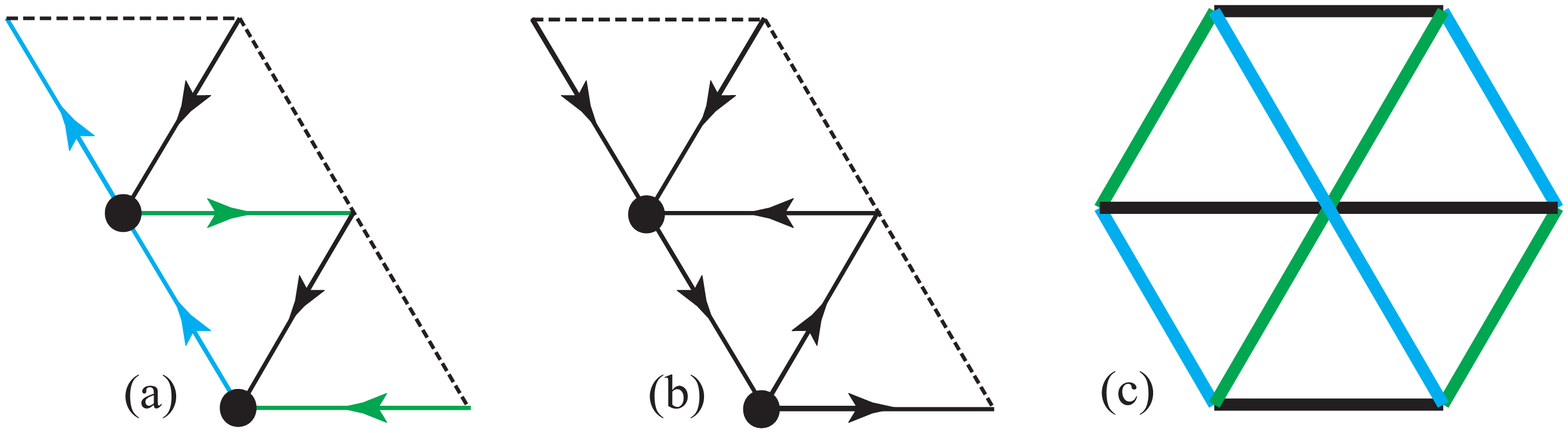}
\caption{ (a) The real space pattern of nearest neighbor (NN) boson pairing amplitude $A_{ij}$, where a direction is assigned due to the fact that $A_{ij}=-A_{ji}$. Black arrows denote phase $\pi/2$, blue arrows denote phase $\pi/6$, and green arrows denote phase $5\pi/6$. (b) The real space pattern of NN boson (fermion) hopping amplitudes $B_{ij}$ ($\chi_{ij}$), where a direction is assigned due to the fact that $B_{ij}=B_{ji}^*$ and  $\chi_{ij}=\chi_{ji}^*$, and the color denotes phase $\pi/2$. The two black dots in (a) and (b) make one unit cell in this PSG ansartz. (c) The real space pairing pattern of the d+id SC order parameter, where black bonds denote pairing $\Delta$, green bonds $\Delta \cdot exp(i 2\pi/3)$, and blue bonds $\Delta \cdot exp(i 4\pi/3)$. \label{pattern}}
\end{figure}

We first construct the c-SDW wave functions using the slave-fermion approach \cite{PhysRevB.74.174423, PhysRevB.45.12377, doi:10.1142/S0217979291000158, PhysRevLett.66.1773, PhysRevB.38.316}, where we rewrite the electron annihilation operator as bosonic spinons and fermionic spinless holons:
\begin{equation}
c_{i\alpha}=b_{i\alpha}f_i^\dagger
\end{equation}
Rewriting the t-J Hamiltonian Eq.(\ref{hamiltonian}) into spinons and holons, the Hamiltonian can be split into two parts at the mean field level: the bosonic part, which describes a bosonic superconductor, and the fermionic part, describing a charge Chern insulator:
\begin{equation}
\begin{split}
H_{c-SDW}^{MF}(b)=& \sum_{ij}( B_{ij}b_{i\alpha}^\dagger b_{i\alpha}+A_{ij}b_{i\alpha}b_{j\beta}\epsilon+h.c. ) \\ & -\mu_b\sum_i b_{i\alpha}^\dagger b_{i\alpha} \\
H_{c-SDW}^{MF}(f)=& \sum_{ij}( \chi_{ij}f_i^\dagger f_j+h.c. )-\mu_f\sum_i f_i^\dagger f_i \\
\end{split}
\end{equation}
where $B_{ij}$ and $A_{ij}$ are the boson singlet hopping and pairing parameters on bond ij, $\chi_{ij}$ is the spinless fermion hopping parameter, $\mu_b$ and $\mu_f$ are the boson and fermion chemical potential respectively. By gluing the wave functions from these two Hamiltonians, one can obtain the wave function describing the whole Hamiltonian. Changing parameters $B_{ij}$, $A_{ij}$, and $\chi_{ij}$, and following the projective symmetry group (PSG) \cite{13,14,15} analysis (see Appendix B), we find the real space pattern of $B_{ij}$, $A_{ij}$ and $\chi_{ij}$ that describes the c-SDW phase, as depicted in Fig.\ref{pattern}(a) and Fig.\ref{pattern}(b). Analogous states with doubled unit cells are called $\pi$ flux states in quantum spin liquids.

The construction of the wave function that describes d+id SC is simpler: we use the slave boson approach \cite{13, 14}, where the electron is split into fermionic spinon and bosonic holon:
\begin{equation}
c_{i\alpha}=f_{i\alpha}b_i^\dagger
\end{equation}
If the fermionic spinons form a d+id band structure while the bosons are condensed at $\Gamma$ point of the Brillouin Zone, then the system gives a d+id superconductor, and the mean field Hamiltonian can be written as:
\begin{equation}
\begin{split}
H_{d+id}^{MF}(f)=& \sum_{ij}( -\chi f_{i\alpha}^\dagger f_{j\alpha}+\Delta_{ij} f_{i\alpha} f_{j\beta}\epsilon_{\alpha \beta}+h.c. ) \\ & -\mu_{f}\sum_{i} f_{i\alpha}^\dagger f_{i\alpha} \\
\end{split}
\end{equation}
where $\chi$ is the hopping parameter, $\delta_{ij}$ are the pairing parameters. The parity of the SC is determined by the symmetry of the pairing parameters. To accommodate d+id SC, we set the relative phases of $\Delta_{ij}$, as depicted in Fig.\ref{pattern}(c), with bonds of different directions being $\Delta$, $\Delta \cdot exp(i 2\pi/3)$ and $\Delta \cdot exp(i 4\pi/3)$.

 \section{\label{sec:level2} Numerical Simulations }
To sharply distinguish candidate phases on finite size samples, we analytically computed the symmetry quantum numbers of c-SDW and d+id SC states which are further confirmed by variational Monte Carlo numerics. Let $\ket{\psi}$ be the many-body state, and $\hat{O}$ be the symmetry operator, then $\hat{O} \ket{\psi}=e^{i\phi}\ket{\psi}$ is the transformed state and $e^{i\phi}$ is the corresponding many-body quantum number. This quantum number can be computed by taking the ratio: $\langle\{s\}|\hat O\ket{\psi}/\langle\{s\}\ket{\psi}=\langle \hat O^{\dagger}\{s\}| \ket{\psi}/\langle\{s\}\ket{\psi}$, where $\{s\}$ is state labeled by a real space spin and hole configuration.

\begin{figure}[t]
\centering
\includegraphics[width=3.4in]{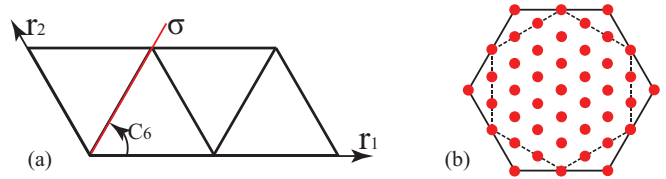}
\caption{  (a) Coordinate system of triangular lattice. (b) k points in Brillouin zone of 36-sites sample.\label{symmetry}}
\end{figure}

We focus on symmetry operators $T_1$, $T_2$, $C_6$ and inversion(i.e. $C_6^3$), where $T_1(T_2)$ is the lattice translation along $\vec{r}_{1(2)}$ and $C_6$ is the $\pi/3$ rotation, as shown in Fig.\ref{symmetry}. As listed in Table \ref{qn}, we computed the ground state quantum numbers of c-SDW and d+id SC on various samples. Between the 16-sites and 36-sites samples (see Fig.\ref{samples_and_pd}(a)) all the considered symmetry quantum numbers are identical, preventing us from distinguishing c-SDW and d+id SC. However, the chosen 28-sites hexagonal sample (see Fig.\ref{samples_and_pd}(b)) is suitable to sharply distinguish these phases. Note that the d+id SC (or c-SDW) phase breaks the time-reversal symmetry and one can construct two wave functions that are time-reversal images of each other. When $C_6$ quantum numbers are different for these two wave functions, they will form a two-fold irreducible representation of the global symmetry.

\begin{figure}[t]
\centering
\includegraphics[width=3.4in]{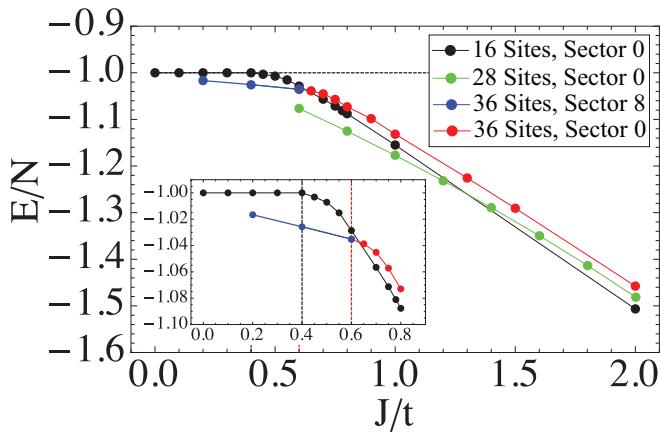}
\caption{ Energy per site of t-J model from DMRG simulations on samples of 16-sites, 28-sites and 36 sites. For 16-sites and 28-sites, the ground states stay in spin sector 0, while for 36-sites, the ground state is in sector 8 for $J/t<0.6$ (blue) and sector 0 for $J/t>0.6$ (red). The horizontal black dashed line indicates the energy per site of the ferromagnetic phase. Inset: Transition region for 16-sites and 36-sites samples. The transition point is $J/t=0.4$ (black vertical dashed line) for 16-sites sample, and $J/t=0.6$ (red vertical dashed line) for 36-sites sample.{\color{red}}\label{dmrg_energy}}
\end{figure}

\begin{table}[t]
\centering
\tabcolsep=10pt
\begin{ruledtabular}

\begin{tabular}{ c  c }
Symmetry	&	c-SDW or d+id SC	\\
\hline
$T_1$	&	$1$	\\
$T_2$	&	$1$	\\
$C_6$	&	$\begin{pmatrix} e^{i2\pi/3}&0\\0&e^{-i2\pi/3} \end{pmatrix}$	\\
Inversion	&	$1$	\\
\end{tabular}

\begin{flushleft}
(a) 16 or 36 sites sample.
\end{flushleft}

\begin{tabular}{ c  c  c }
Symmetry	&	c-SDW	&	d+id SC	\\
\hline
$T_1$	&	$1$	&	$1$	\\
$T_2$	&	$1$	&	$1$	\\
$C_6$	&	$1$	&	$\begin{pmatrix} e^{i2\pi/3}&0\\0&e^{-i2\pi/3} \end{pmatrix}$	\\
Inversion	&	$1$	&	$1$	\\
\end{tabular}

\begin{flushleft}
(b) 28 sites sample.
\end{flushleft}

\end{ruledtabular}

\caption{\label{qn} Variational Monte Carlo results on quantum numbers of c-SDW state and d+id SC state of 16-sites, 36-sites and 28-sites samples. For 16-sites or 36-sites samples, the ground state wave functions of the two candidate phases from two-fold irreducible representations (irreps) of symmetry group with identical symmetry quantum numbers. For 28-sites sample, the ground state wave function of c-SDW forms 1D irreps while that of d+id SC forms 2D irreps.}
\end{table}

\begin{table*}[t]
\centering
\tabcolsep=10pt
\begin{ruledtabular}

\begin{tabular}{ c c c c c c}
$J/t$		&	$\Delta_1$	&	$\Delta_2$	&	$\Delta_3$	&	$\frac{Arg(\Delta_2)-Arg(\Delta_1)}{2\pi}$		&	$\frac{Arg(\Delta_3)-Arg(\Delta_1)}{2\pi}$		\\
\hline
0.6	&	0.001783	&0.001287	&0.001837	&-0.277946	&0.286516\\
0.8	&	0.003922	&0.002356	&0.002044	&-0.268602	&0.340033\\
1.0	&	0.003117	&0.002885	&0.002801	&-0.354765	&0.328343\\
1.2	&	0.004177	&0.003216	&0.003289	&-0.344315	&0.347355\\
1.4	&	0.005628	&0.004387	&0.004712	&-0.347967	&0.358649\\
1.6	&	0.008756	&0.006370	&0.006951	&-0.348524	&0.371596\\
1.8	&	0.013474	&0.010153	&0.009597	&-0.369925	&0.352528\\
2.0	&	0.017143	&0.012216	&0.011950	&-0.367776	&0.366071\\

\end{tabular}
\end{ruledtabular}

\caption{Pair-pair correlation function in DMRG of t-J model on 28-sites sample.The correlation function $\Delta_\alpha=\left< \hat{B}_{ij}^\dagger \hat{B}_{kl} \right>$ is chosen as nearest neighbor bonds with farthest distance in the sample, while keeping bonds $ij$ fixed as the solid bond in Fig.\ref{samples_and_pd}(b) and bonds $kl$ the three dashed bonds, where $\alpha=1,2,3$ are the specific bond $kl$ indices relabeled as in Fig.\ref{samples_and_pd}(b). \label{pairpair} }
\end{table*}

We perform DMRG simulations for the three samples in Fig.\ref{samples_and_pd}. First, the ground state energy has already provided useful information, see Fig.\ref{dmrg_energy}. For 16-sites sample, there is a horizontal plateau in the region of $0\le J/t<0.4$ whose energy stays at -1. This is a signature of the ferromagnetic order. Because with spin-polarized electrons, the J term in t-J model vanishes and the ground state energy can be computed based on the non-interacting hopping problem. On this sample, by filling all the 8 electrons in the spin-polarized band, one finds the energy per site to be $E/N=-2\times8/16=-1$. For the 36-sites sample with 18 electrons, the situation can be understood as follows. We label the sectors by the total $S_z$ spin quantum number. There are 10 non-negative $S_z$ sectors ranging from 0 to 9, with $S_z=9$ being fully polarized along the $z$-direction. The fully polarized ferromagnetic state still produces $E/N=-1$, but it is not the ground state. Instead, the ground states are found to form a total spin $S=8$ representation and one of them is in the $S_z=8$ sector, producing an energy curve with a slight slope as indicated by the blue curve in Fig.\ref{dmrg_energy}. We believe that this result is a consequence of the specific energy shell structure on this sample: Let us denote the majority spin flavor to be spin up, then 17 spin-up electrons would fully fill the energy shell for the spinless hopping Hamiltonian (Fig.\ref{symmetry}(b)), leaving one extra down spin. It is reasonable to expect that this is a finite size artifact and  the fully polarized ferromagnetism would be restored in the thermodynamic limit.

\begin{figure}[t]
\centering
\includegraphics[width=3.4in]{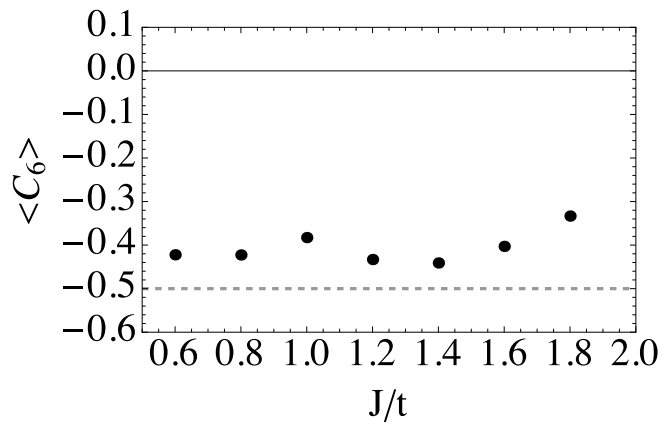}
\caption{ Quantum numbers of $C_6$ in DMRG ground state for t-J model on 28-sites sample with projection of wave function to center of mass at $\Gamma$ point in the Brillouin zone. \label{c6}}
\end{figure}

To understand the nature of the phase with large $J/t$, we calculate the symmetry quantum numbers of the DMRG ground states. We find $\left< T_1 \right>=1$, $\left< T_2 \right>=1$ and $\left< Inv \right>=1$ for the ground states in this regime on all the three samples, which is consistent with both candidate phases (see Table \ref{qn}). However, the $C_6$ quantum number of DMRG ground state on the 28-sites sample can be used to sharply distinguish the phases, and we find it to be consistent with the d+id SC phase. Technically, here we take advantage of the fact that the model Hamiltonian is purely real (in the real-space spin configuration basis). Consequently the DMRG simulation gives a purely real ground state wave function $|\psi_{dmrg}\rangle$. If the ground states of the model form a two-fold irreducible representation as the d+id SC on the 28-sites sample(see Table \ref{qn}), then the DMRG wave function would be an equal weight superposition of the $C_6=e^{i2\pi/3}$ and $C_6=e^{-i2\pi/3}$ states, and the expectation value of the $C_6$ transformation operator: $\langle \psi_{dmrg}|\hat C_6|\psi_{dmrg}\rangle$ would be $-\frac{1}{2}$. On the other hand, this expectation value would be $1$ for the c-SDW state. We have measured this expectation value in the large $J/t$ regime using the standard Monte Carlo technique(see Fig.\ref{c6}), and the result is clearly consistent with $-\frac{1}{2}$. (Here we have projected the $|\psi_{dmrg}\rangle$ to the center of momentum $\Gamma=(0,0)$ point for better convergence.)

\begin{figure}[t]
\centering
\includegraphics[width=3.4in]{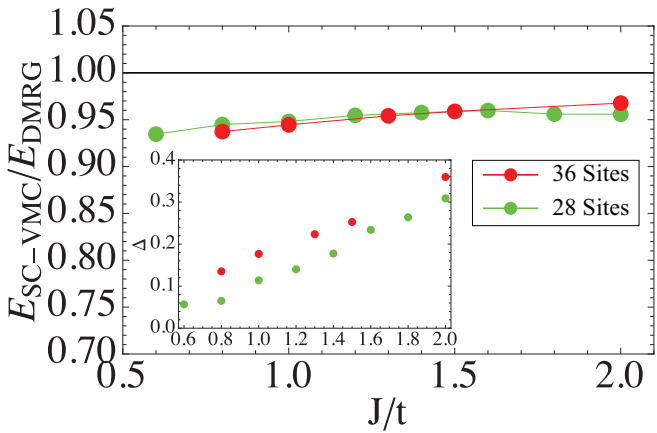}
\caption{ The minimal energy of the Gutzwiller projected d+id variational wave function, varying the single mean field parameter $\Delta$. Inset: The optimal value of $\Delta$ that minimizes the ground state energy. \label{energy_compare}}
\end{figure} 

Next we confirm the nature of the DMRG ground state in two more ways: (1) The energy of the d+id SC trial wave function proposed in Section II is calculated using the variational Monte Carlo technique\cite{10-2} and compared to the DMRG ground state. As can be seen in Fig.\ref{energy_compare}, the optimal d+id SC wave function obtained by tuning only one variational parameter $\Delta$ (the nearest neighbor pairing) can already produce $\sim95\%$ of the energy of DMRG ground state energy. (2) The pair-pair correlation function of the DMRG ground state are measured, as shown Table \ref{pairpair}. One key signature of the d+id pairing is the relative phase of the pairing on the real space bonds, as depicted in Fig.\ref{pattern}(c). We calculated the pair-pair correlation function, defined as $\left< \hat{B}_{ij}^\dagger \hat{B}_{kl} \right>$, where $\hat{B}_{ij}=c_{i \uparrow} c_{j \downarrow}-c_{i \downarrow} c_{j \uparrow}$ is the singlet pairing, while $i$ and $j$ are two nearest neighbor sites forming a bond. The bonds are chosen with farthest distance in the sample, with $ij$ fixed and $kl$ in three options varied by directions, as shown in Fig.\ref{samples_and_pd}(b). We find that the pattern of the relative pairing phases is consistent with the d+id SC.


{\section{\label{sec:level3} Discussion and Conclusion}} 

Based on the idea of using quantum numbers in finite system sizes to sharply distinguish candidate quantum phases, we investigate the quantum phase diagram of the t-J model on the triangular lattice at 1/2 doping. As shown in Fig.\ref{samples_and_pd}, ferromagnetic phase is realized in the the small $J/t$ regime, separated from a chiral d+id superconductor phase in the large $J/t$ regime.

The crucial advantage of the method adopted here is the capability to sharply distinguish candidate quantum phases even on small to intermediate size samples. To achieve this goal, we purposely choose samples so that candidate quantum phases have distinct quantum numbers. For instance, the c-SDW phase and the d+id superconductor share the same quantum numbers on the 36-sites sample, but feature different lattice quantum numbers on the 28-sites sample. Such analytical understanding of the symmetry properties provides important guidance in our numerical simulations. 

One may wonder about the generality of this method. Indeed, in this study, we are fortunate to be able to only focus on a small number of candidate quantum phases, based on the special shape of the non-interacting Fermi surface and previous weak coupling analysis \cite{7}. But in a general model, in principle one needs to take into account a large number of candidate phases and the present treatment scheme may become intractable. In this regard, hopefully one could develop a theoretical method to systematically diagnose symmetry properties of the quantum wave functions of candidate phases. A recent work based on tensor-network formulation is an attempt to develop such a general method \cite{16}.

This work is supported by Alfred P. Sloan fellowship and National Science Foundation under Grant No. DMR-1151440.

\bibliographystyle{apsrev}
\bibliography{reference}

\begin{thebibliography}{38}
\expandafter\ifx\csname natexlab\endcsname\relax\def\natexlab#1{#1}\fi
\expandafter\ifx\csname bibnamefont\endcsname\relax
  \def\bibnamefont#1{#1}\fi
\expandafter\ifx\csname bibfnamefont\endcsname\relax
  \def\bibfnamefont#1{#1}\fi
\expandafter\ifx\csname citenamefont\endcsname\relax
  \def\citenamefont#1{#1}\fi
\expandafter\ifx\csname url\endcsname\relax
  \def\url#1{\texttt{#1}}\fi
\expandafter\ifx\csname urlprefix\endcsname\relax\def\urlprefix{URL }\fi
\providecommand{\bibinfo}[2]{#2}
\providecommand{\eprint}[2][]{\url{#2}}

\bibitem[{\citenamefont{Wang et~al.}(2004)\citenamefont{Wang, Lee, and
  Lee}}]{1}
\bibinfo{author}{\bibfnamefont{Q.-H.} \bibnamefont{Wang}},
  \bibinfo{author}{\bibfnamefont{D.-H.} \bibnamefont{Lee}}, \bibnamefont{and}
  \bibinfo{author}{\bibfnamefont{P.~A.} \bibnamefont{Lee}},
  \bibinfo{journal}{Phys. Rev. B} \textbf{\bibinfo{volume}{69}},
  \bibinfo{pages}{092504} (\bibinfo{year}{2004}).

\bibitem[{\citenamefont{Ralko et~al.}(2015)\citenamefont{Ralko, Merino, and
  Fratini}}]{PhysRevB.91.165139}
\bibinfo{author}{\bibfnamefont{A.}~\bibnamefont{Ralko}},
  \bibinfo{author}{\bibfnamefont{J.}~\bibnamefont{Merino}}, \bibnamefont{and}
  \bibinfo{author}{\bibfnamefont{S.}~\bibnamefont{Fratini}},
  \bibinfo{journal}{Phys. Rev. B} \textbf{\bibinfo{volume}{91}},
  \bibinfo{pages}{165139} (\bibinfo{year}{2015}).

\bibitem[{\citenamefont{Becker et~al.}(2015)\citenamefont{Becker, Hermanns,
  Bauer, Garst, and Trebst}}]{PhysRevB.91.155135}
\bibinfo{author}{\bibfnamefont{M.}~\bibnamefont{Becker}},
  \bibinfo{author}{\bibfnamefont{M.}~\bibnamefont{Hermanns}},
  \bibinfo{author}{\bibfnamefont{B.}~\bibnamefont{Bauer}},
  \bibinfo{author}{\bibfnamefont{M.}~\bibnamefont{Garst}}, \bibnamefont{and}
  \bibinfo{author}{\bibfnamefont{S.}~\bibnamefont{Trebst}},
  \bibinfo{journal}{Phys. Rev. B} \textbf{\bibinfo{volume}{91}},
  \bibinfo{pages}{155135} (\bibinfo{year}{2015}).

\bibitem[{\citenamefont{Pen et~al.}(1997)\citenamefont{Pen, van~den Brink,
  Khomskii, and Sawatzky}}]{PhysRevLett.78.1323}
\bibinfo{author}{\bibfnamefont{H.~F.} \bibnamefont{Pen}},
  \bibinfo{author}{\bibfnamefont{J.}~\bibnamefont{van~den Brink}},
  \bibinfo{author}{\bibfnamefont{D.~I.} \bibnamefont{Khomskii}},
  \bibnamefont{and} \bibinfo{author}{\bibfnamefont{G.~A.}
  \bibnamefont{Sawatzky}}, \bibinfo{journal}{Phys. Rev. Lett.}
  \textbf{\bibinfo{volume}{78}}, \bibinfo{pages}{1323} (\bibinfo{year}{1997}).

\bibitem[{\citenamefont{Shimizu et~al.}(2003)\citenamefont{Shimizu, Miyagawa,
  Kanoda, Maesato, and Saito}}]{2-1}
\bibinfo{author}{\bibfnamefont{Y.}~\bibnamefont{Shimizu}},
  \bibinfo{author}{\bibfnamefont{K.}~\bibnamefont{Miyagawa}},
  \bibinfo{author}{\bibfnamefont{K.}~\bibnamefont{Kanoda}},
  \bibinfo{author}{\bibfnamefont{M.}~\bibnamefont{Maesato}}, \bibnamefont{and}
  \bibinfo{author}{\bibfnamefont{G.}~\bibnamefont{Saito}},
  \bibinfo{journal}{Phys. Rev. Lett.} \textbf{\bibinfo{volume}{91}},
  \bibinfo{pages}{107001} (\bibinfo{year}{2003}).

\bibitem[{\citenamefont{Itou et~al.}(2008)\citenamefont{Itou, Oyamada, Maegawa,
  Tamura, and Kato}}]{2-2}
\bibinfo{author}{\bibfnamefont{T.}~\bibnamefont{Itou}},
  \bibinfo{author}{\bibfnamefont{A.}~\bibnamefont{Oyamada}},
  \bibinfo{author}{\bibfnamefont{S.}~\bibnamefont{Maegawa}},
  \bibinfo{author}{\bibfnamefont{M.}~\bibnamefont{Tamura}}, \bibnamefont{and}
  \bibinfo{author}{\bibfnamefont{R.}~\bibnamefont{Kato}},
  \bibinfo{journal}{Phys. Rev. B} \textbf{\bibinfo{volume}{77}},
  \bibinfo{pages}{104413} (\bibinfo{year}{2008}).

\bibitem[{\citenamefont{Yamashita et~al.}(2008)\citenamefont{Yamashita,
  Nakazawa, Oguni, Oshima, Nojiri, Shimizu, Miyagawa, and
  Kanoda}}]{Yamashita2008}
\bibinfo{author}{\bibfnamefont{S.}~\bibnamefont{Yamashita}},
  \bibinfo{author}{\bibfnamefont{Y.}~\bibnamefont{Nakazawa}},
  \bibinfo{author}{\bibfnamefont{M.}~\bibnamefont{Oguni}},
  \bibinfo{author}{\bibfnamefont{Y.}~\bibnamefont{Oshima}},
  \bibinfo{author}{\bibfnamefont{H.}~\bibnamefont{Nojiri}},
  \bibinfo{author}{\bibfnamefont{Y.}~\bibnamefont{Shimizu}},
  \bibinfo{author}{\bibfnamefont{K.}~\bibnamefont{Miyagawa}}, \bibnamefont{and}
  \bibinfo{author}{\bibfnamefont{K.}~\bibnamefont{Kanoda}},
  \bibinfo{journal}{Nat Phys} \textbf{\bibinfo{volume}{4}},
  \bibinfo{pages}{459} (\bibinfo{year}{2008}).

\bibitem[{\citenamefont{Kino and Kontani}(1998)}]{doi:10.1143/JPSJ.67.3691}
\bibinfo{author}{\bibfnamefont{H.}~\bibnamefont{Kino}} \bibnamefont{and}
  \bibinfo{author}{\bibfnamefont{H.}~\bibnamefont{Kontani}},
  \bibinfo{journal}{Journal of the Physical Society of Japan}
  \textbf{\bibinfo{volume}{67}}, \bibinfo{pages}{3691} (\bibinfo{year}{1998}).

\bibitem[{\citenamefont{Kimura et~al.}(2006)\citenamefont{Kimura, Lashley, and
  Ramirez}}]{3-1}
\bibinfo{author}{\bibfnamefont{T.}~\bibnamefont{Kimura}},
  \bibinfo{author}{\bibfnamefont{J.~C.} \bibnamefont{Lashley}},
  \bibnamefont{and} \bibinfo{author}{\bibfnamefont{A.~P.}
  \bibnamefont{Ramirez}}, \bibinfo{journal}{Phys. Rev. B}
  \textbf{\bibinfo{volume}{73}}, \bibinfo{pages}{220401}
  (\bibinfo{year}{2006}).

\bibitem[{\citenamefont{Chung et~al.}(2001)\citenamefont{Chung, Marston, and
  McKenzie}}]{3-2}
\bibinfo{author}{\bibfnamefont{C.~H.} \bibnamefont{Chung}},
  \bibinfo{author}{\bibfnamefont{J.~B.} \bibnamefont{Marston}},
  \bibnamefont{and} \bibinfo{author}{\bibfnamefont{R.~H.}
  \bibnamefont{McKenzie}}, \bibinfo{journal}{Journal of Physics: Condensed
  Matter} \textbf{\bibinfo{volume}{13}}, \bibinfo{pages}{5159}
  (\bibinfo{year}{2001}).

\bibitem[{\citenamefont{Schaak et~al.}(2003)\citenamefont{Schaak, Klimczuk,
  Foo, and Cava}}]{4}
\bibinfo{author}{\bibfnamefont{R.}~\bibnamefont{Schaak}},
  \bibinfo{author}{\bibfnamefont{T.}~\bibnamefont{Klimczuk}},
  \bibinfo{author}{\bibfnamefont{M.~L.} \bibnamefont{Foo}}, \bibnamefont{and}
  \bibinfo{author}{\bibfnamefont{R.~J.} \bibnamefont{Cava}},
  \bibinfo{journal}{Nature} \textbf{\bibinfo{volume}{424}},
  \bibinfo{pages}{527} (\bibinfo{year}{2003}).

\bibitem[{\citenamefont{Kumar and Shastry}(2003)}]{PhysRevB.68.104508}
\bibinfo{author}{\bibfnamefont{B.}~\bibnamefont{Kumar}} \bibnamefont{and}
  \bibinfo{author}{\bibfnamefont{B.~S.} \bibnamefont{Shastry}},
  \bibinfo{journal}{Phys. Rev. B} \textbf{\bibinfo{volume}{68}},
  \bibinfo{pages}{104508} (\bibinfo{year}{2003}).

\bibitem[{\citenamefont{Li et~al.}(2015)\citenamefont{Li, Chen, Tong, Pi, Liu,
  Yang, Wang, and Zhang}}]{5}
\bibinfo{author}{\bibfnamefont{Y.}~\bibnamefont{Li}},
  \bibinfo{author}{\bibfnamefont{G.}~\bibnamefont{Chen}},
  \bibinfo{author}{\bibfnamefont{W.}~\bibnamefont{Tong}},
  \bibinfo{author}{\bibfnamefont{L.}~\bibnamefont{Pi}},
  \bibinfo{author}{\bibfnamefont{J.}~\bibnamefont{Liu}},
  \bibinfo{author}{\bibfnamefont{Z.}~\bibnamefont{Yang}},
  \bibinfo{author}{\bibfnamefont{X.}~\bibnamefont{Wang}}, \bibnamefont{and}
  \bibinfo{author}{\bibfnamefont{Q.}~\bibnamefont{Zhang}},
  \bibinfo{journal}{Phys. Rev. Lett.} \textbf{\bibinfo{volume}{115}},
  \bibinfo{pages}{167203} (\bibinfo{year}{2015}).

\bibitem[{\citenamefont{Yamashita et~al.}(2009)\citenamefont{Yamashita, Nakata,
  Kasahara, Sasaki, Yoneyama, Kobayashi, Fujimoto, Shibauchi, and
  Matsuda}}]{Yamashita2009}
\bibinfo{author}{\bibfnamefont{M.}~\bibnamefont{Yamashita}},
  \bibinfo{author}{\bibfnamefont{N.}~\bibnamefont{Nakata}},
  \bibinfo{author}{\bibfnamefont{Y.}~\bibnamefont{Kasahara}},
  \bibinfo{author}{\bibfnamefont{T.}~\bibnamefont{Sasaki}},
  \bibinfo{author}{\bibfnamefont{N.}~\bibnamefont{Yoneyama}},
  \bibinfo{author}{\bibfnamefont{N.}~\bibnamefont{Kobayashi}},
  \bibinfo{author}{\bibfnamefont{S.}~\bibnamefont{Fujimoto}},
  \bibinfo{author}{\bibfnamefont{T.}~\bibnamefont{Shibauchi}},
  \bibnamefont{and} \bibinfo{author}{\bibfnamefont{Y.}~\bibnamefont{Matsuda}},
  \bibinfo{journal}{Nat Phys} \textbf{\bibinfo{volume}{5}}, \bibinfo{pages}{44}
  (\bibinfo{year}{2009}).

\bibitem[{\citenamefont{Eisaki et~al.}(1994)\citenamefont{Eisaki, Takagi, Cava,
  Batlogg, Krajewski, Peck, Mizuhashi, Lee, and Uchida}}]{PhysRevB.50.647}
\bibinfo{author}{\bibfnamefont{H.}~\bibnamefont{Eisaki}},
  \bibinfo{author}{\bibfnamefont{H.}~\bibnamefont{Takagi}},
  \bibinfo{author}{\bibfnamefont{R.~J.} \bibnamefont{Cava}},
  \bibinfo{author}{\bibfnamefont{B.}~\bibnamefont{Batlogg}},
  \bibinfo{author}{\bibfnamefont{J.~J.} \bibnamefont{Krajewski}},
  \bibinfo{author}{\bibfnamefont{W.~F.} \bibnamefont{Peck}},
  \bibinfo{author}{\bibfnamefont{K.}~\bibnamefont{Mizuhashi}},
  \bibinfo{author}{\bibfnamefont{J.~O.} \bibnamefont{Lee}}, \bibnamefont{and}
  \bibinfo{author}{\bibfnamefont{S.}~\bibnamefont{Uchida}},
  \bibinfo{journal}{Phys. Rev. B} \textbf{\bibinfo{volume}{50}},
  \bibinfo{pages}{647} (\bibinfo{year}{1994}).

\bibitem[{\citenamefont{Jiang et~al.}(2015)\citenamefont{Jiang, Zhang, Zhou,
  and Wang}}]{PhysRevLett.114.216402}
\bibinfo{author}{\bibfnamefont{K.}~\bibnamefont{Jiang}},
  \bibinfo{author}{\bibfnamefont{Y.}~\bibnamefont{Zhang}},
  \bibinfo{author}{\bibfnamefont{S.}~\bibnamefont{Zhou}}, \bibnamefont{and}
  \bibinfo{author}{\bibfnamefont{Z.}~\bibnamefont{Wang}},
  \bibinfo{journal}{Phys. Rev. Lett.} \textbf{\bibinfo{volume}{114}},
  \bibinfo{pages}{216402} (\bibinfo{year}{2015}).

\bibitem[{\citenamefont{Martin and Batista}(2008)}]{6}
\bibinfo{author}{\bibfnamefont{I.}~\bibnamefont{Martin}} \bibnamefont{and}
  \bibinfo{author}{\bibfnamefont{C.~D.} \bibnamefont{Batista}},
  \bibinfo{journal}{Phys. Rev. Lett.} \textbf{\bibinfo{volume}{101}},
  \bibinfo{pages}{156402} (\bibinfo{year}{2008}).

\bibitem[{\citenamefont{Nandkishore et~al.}(2014)\citenamefont{Nandkishore,
  Thomale, and Chubukov}}]{7}
\bibinfo{author}{\bibfnamefont{R.}~\bibnamefont{Nandkishore}},
  \bibinfo{author}{\bibfnamefont{R.}~\bibnamefont{Thomale}}, \bibnamefont{and}
  \bibinfo{author}{\bibfnamefont{A.~V.} \bibnamefont{Chubukov}},
  \bibinfo{journal}{Phys. Rev. B} \textbf{\bibinfo{volume}{89}},
  \bibinfo{pages}{144501} (\bibinfo{year}{2014}).

\bibitem[{\citenamefont{Nagaoka}(1966)}]{8}
\bibinfo{author}{\bibfnamefont{Y.}~\bibnamefont{Nagaoka}},
  \bibinfo{journal}{Phys. Rev.} \textbf{\bibinfo{volume}{147}},
  \bibinfo{pages}{392} (\bibinfo{year}{1966}).

\bibitem[{\citenamefont{Schollw\"ock}(2005)}]{RevModPhys.77.259}
\bibinfo{author}{\bibfnamefont{U.}~\bibnamefont{Schollw\"ock}},
  \bibinfo{journal}{Rev. Mod. Phys.} \textbf{\bibinfo{volume}{77}},
  \bibinfo{pages}{259} (\bibinfo{year}{2005}).

\bibitem[{\citenamefont{Schollw{\"o}ck}(2011)}]{9-1}
\bibinfo{author}{\bibfnamefont{U.}~\bibnamefont{Schollw{\"o}ck}},
  \bibinfo{journal}{Annals of Physics} \textbf{\bibinfo{volume}{326}},
  \bibinfo{pages}{96} (\bibinfo{year}{2011}).

\bibitem[{\citenamefont{White}(1992)}]{9-2}
\bibinfo{author}{\bibfnamefont{S.~R.} \bibnamefont{White}},
  \bibinfo{journal}{Physical Review Letters} \textbf{\bibinfo{volume}{69}},
  \bibinfo{pages}{2863} (\bibinfo{year}{1992}).

\bibitem[{\citenamefont{Hallberg}(2006)}]{doi:10.1080/00018730600766432}
\bibinfo{author}{\bibfnamefont{K.~A.} \bibnamefont{Hallberg}},
  \bibinfo{journal}{Advances in Physics} \textbf{\bibinfo{volume}{55}},
  \bibinfo{pages}{477} (\bibinfo{year}{2006}).

\bibitem[{\citenamefont{Foulkes et~al.}(2001)\citenamefont{Foulkes, Mitas,
  Needs, and Rajagopal}}]{10-1}
\bibinfo{author}{\bibfnamefont{W.}~\bibnamefont{Foulkes}},
  \bibinfo{author}{\bibfnamefont{L.}~\bibnamefont{Mitas}},
  \bibinfo{author}{\bibfnamefont{R.}~\bibnamefont{Needs}}, \bibnamefont{and}
  \bibinfo{author}{\bibfnamefont{G.}~\bibnamefont{Rajagopal}},
  \bibinfo{journal}{Reviews of Modern Physics} \textbf{\bibinfo{volume}{73}},
  \bibinfo{pages}{33} (\bibinfo{year}{2001}).

\bibitem[{\citenamefont{Gros}(1989)}]{10-2}
\bibinfo{author}{\bibfnamefont{C.}~\bibnamefont{Gros}},
  \bibinfo{journal}{Annals of Physics} \textbf{\bibinfo{volume}{189}},
  \bibinfo{pages}{53} (\bibinfo{year}{1989}).

\bibitem[{\citenamefont{Ceperley et~al.}(1977)\citenamefont{Ceperley, Chester,
  and Kalos}}]{PhysRevB.16.3081}
\bibinfo{author}{\bibfnamefont{D.}~\bibnamefont{Ceperley}},
  \bibinfo{author}{\bibfnamefont{G.~V.} \bibnamefont{Chester}},
  \bibnamefont{and} \bibinfo{author}{\bibfnamefont{M.~H.} \bibnamefont{Kalos}},
  \bibinfo{journal}{Phys. Rev. B} \textbf{\bibinfo{volume}{16}},
  \bibinfo{pages}{3081} (\bibinfo{year}{1977}).

\bibitem[{\citenamefont{Harju et~al.}(1997)\citenamefont{Harju, Barbiellini,
  Siljam\"aki, Nieminen, and Ortiz}}]{PhysRevLett.79.1173}
\bibinfo{author}{\bibfnamefont{A.}~\bibnamefont{Harju}},
  \bibinfo{author}{\bibfnamefont{B.}~\bibnamefont{Barbiellini}},
  \bibinfo{author}{\bibfnamefont{S.}~\bibnamefont{Siljam\"aki}},
  \bibinfo{author}{\bibfnamefont{R.~M.} \bibnamefont{Nieminen}},
  \bibnamefont{and} \bibinfo{author}{\bibfnamefont{G.}~\bibnamefont{Ortiz}},
  \bibinfo{journal}{Phys. Rev. Lett.} \textbf{\bibinfo{volume}{79}},
  \bibinfo{pages}{1173} (\bibinfo{year}{1997}).

\bibitem[{\citenamefont{Jiang et~al.}(2014)\citenamefont{Jiang, Mesaros, and
  Ran}}]{11}
\bibinfo{author}{\bibfnamefont{S.}~\bibnamefont{Jiang}},
  \bibinfo{author}{\bibfnamefont{A.}~\bibnamefont{Mesaros}}, \bibnamefont{and}
  \bibinfo{author}{\bibfnamefont{Y.}~\bibnamefont{Ran}},
  \bibinfo{journal}{Physical Review X} \textbf{\bibinfo{volume}{4}},
  \bibinfo{pages}{031040} (\bibinfo{year}{2014}).

\bibitem[{\citenamefont{Jeckelmann and White}(1998)}]{12}
\bibinfo{author}{\bibfnamefont{E.}~\bibnamefont{Jeckelmann}} \bibnamefont{and}
  \bibinfo{author}{\bibfnamefont{S.~R.} \bibnamefont{White}},
  \bibinfo{journal}{Phys. Rev. B} \textbf{\bibinfo{volume}{57}},
  \bibinfo{pages}{6376} (\bibinfo{year}{1998}).

\bibitem[{\citenamefont{Wang and Vishwanath}(2006)}]{PhysRevB.74.174423}
\bibinfo{author}{\bibfnamefont{F.}~\bibnamefont{Wang}} \bibnamefont{and}
  \bibinfo{author}{\bibfnamefont{A.}~\bibnamefont{Vishwanath}},
  \bibinfo{journal}{Phys. Rev. B} \textbf{\bibinfo{volume}{74}},
  \bibinfo{pages}{174423} (\bibinfo{year}{2006}).

\bibitem[{\citenamefont{Sachdev}(1992)}]{PhysRevB.45.12377}
\bibinfo{author}{\bibfnamefont{S.}~\bibnamefont{Sachdev}},
  \bibinfo{journal}{Phys. Rev. B} \textbf{\bibinfo{volume}{45}},
  \bibinfo{pages}{12377} (\bibinfo{year}{1992}).

\bibitem[{\citenamefont{Sachdev and
  Read}(1991)}]{doi:10.1142/S0217979291000158}
\bibinfo{author}{\bibfnamefont{S.}~\bibnamefont{Sachdev}} \bibnamefont{and}
  \bibinfo{author}{\bibfnamefont{N.}~\bibnamefont{Read}},
  \bibinfo{journal}{International Journal of Modern Physics B}
  \textbf{\bibinfo{volume}{05}}, \bibinfo{pages}{219} (\bibinfo{year}{1991}).

\bibitem[{\citenamefont{Read and Sachdev}(1991)}]{PhysRevLett.66.1773}
\bibinfo{author}{\bibfnamefont{N.}~\bibnamefont{Read}} \bibnamefont{and}
  \bibinfo{author}{\bibfnamefont{S.}~\bibnamefont{Sachdev}},
  \bibinfo{journal}{Phys. Rev. Lett.} \textbf{\bibinfo{volume}{66}},
  \bibinfo{pages}{1773} (\bibinfo{year}{1991}).

\bibitem[{\citenamefont{Arovas and Auerbach}(1988)}]{PhysRevB.38.316}
\bibinfo{author}{\bibfnamefont{D.~P.} \bibnamefont{Arovas}} \bibnamefont{and}
  \bibinfo{author}{\bibfnamefont{A.}~\bibnamefont{Auerbach}},
  \bibinfo{journal}{Phys. Rev. B} \textbf{\bibinfo{volume}{38}},
  \bibinfo{pages}{316} (\bibinfo{year}{1988}).

\bibitem[{\citenamefont{Wen}(2002{\natexlab{a}})}]{13}
\bibinfo{author}{\bibfnamefont{X.-G.} \bibnamefont{Wen}},
  \bibinfo{journal}{Phys. Rev. B} \textbf{\bibinfo{volume}{65}},
  \bibinfo{pages}{165113} (\bibinfo{year}{2002}{\natexlab{a}}).

\bibitem[{\citenamefont{Wen}(2002{\natexlab{b}})}]{14}
\bibinfo{author}{\bibfnamefont{X.-G.} \bibnamefont{Wen}},
  \bibinfo{journal}{Physics Letters A} \textbf{\bibinfo{volume}{300}},
  \bibinfo{pages}{175 } (\bibinfo{year}{2002}{\natexlab{b}}), ISSN
  \bibinfo{issn}{0375-9601}.

\bibitem[{\citenamefont{Cincio and Vidal}(2013)}]{15}
\bibinfo{author}{\bibfnamefont{L.}~\bibnamefont{Cincio}} \bibnamefont{and}
  \bibinfo{author}{\bibfnamefont{G.}~\bibnamefont{Vidal}},
  \bibinfo{journal}{Phys. Rev. Lett.} \textbf{\bibinfo{volume}{110}},
  \bibinfo{pages}{067208} (\bibinfo{year}{2013}).

\bibitem[{\citenamefont{Jiang and Ran}(2015)}]{16}
\bibinfo{author}{\bibfnamefont{S.}~\bibnamefont{Jiang}} \bibnamefont{and}
  \bibinfo{author}{\bibfnamefont{Y.}~\bibnamefont{Ran}},
  \bibinfo{journal}{Phys. Rev. B} \textbf{\bibinfo{volume}{92}},
  \bibinfo{pages}{104414} (\bibinfo{year}{2015}).

\end{thebibliography}

\newpage
\appendix
\section{\label{sec:level4} Symmetry group of triangular lattice}
Fig.\ref{symmetry}(a) shows the coordinate system we used, and the lattice sites can be labeled by $(x, y)$ where the position of the lattice point $\vec{r}=x \vec{a}_1+y \vec{a}_2$, and $\vec{a}_{1(2)}$ are the Bravais lattice vectors along $\textbf{r}_{1(2)}$. We choose the generator operators of the symmetry group of the triangular lattice as follows: translations $T_{1(2)}$along directions $\textbf{r}_{1(2)}$ by one lattice spacing, a $\pi/3$ rotation $C_6$ in the 2D lattice plane with the rotation center at (0,0) and a mirror reflection with time reversal operation labeled by $\bar{\sigma}$. Under symmetry operations, we find that coordinates transform in the following way:

\begin{equation}
\label{coordinates}
\begin{split}
T_1&: (x,y)\rightarrow(x+1,y)	\\
T_2&: (x,y)\rightarrow(x,y+1)	\\
\bar{\sigma}&: (x,y)\rightarrow(y,x) \\
C_6&: (x,y)\rightarrow(x-y,x) \\
\end{split}
\end{equation}
and the multiplication rules of the symmetry group are determined by:

\begin{equation}
\label{multiplication}
\begin{split}
T_1^{-1}	C_6	T_2^{-1}	C_6^{-1}	&=\textbf{e} \\
T_2^{-1}	C_6	T_1	T_2	C_6^{-1}	&=\textbf{e} \\
T_1^{-1}	\bar{\sigma}	T_2	\bar{\sigma}^{-1}	&=\textbf{e} \\
T_2^{-1}	\bar{\sigma}	T_1	\bar{\sigma}^{-1}	&=\textbf{e} \\
C_6^6=\bar{\sigma}^2	&=\textbf{e} \\
\bar{\sigma}	C_6	\bar{\sigma}	C_6	&=\textbf{e} \\
\end{split}
\end{equation}
where $\textbf{e}$ represents the identity of the symmetry group.
\section{\label{sec:level5} Projective symmetry group (PSG) analysis}
The projective symmetry group (PSG) method is used to classify different mean field ansatze. We associate a $U(1)$ gauge group element $exp(i \phi_X(j))$ to each lattice symmetry group element $X$. Let the mean field ansatz be invariant under:
\begin{equation}
\begin{split}
& A_{X(i)X(j)}=e^{i (\phi_{X(i)}+\phi_{X(j)})} A_{ij} \\
& B_{X(i)X(j)}=e^{-i (\phi_{X(i)}-\phi_{X(j)})} B_{ij} \\
& \chi_{X(i)X(j)}=e^{-i (\phi_{X(i)}-\phi_{X(j)})} \chi_{ij}, \\
\end{split}
\end{equation}
caused by PSG operation that transforms $b_{j\alpha}$ and $f_j$ by a $U(1)$ phase:
\begin{equation}
\begin{split}
& b_{j\alpha} \rightarrow e^{i\phi_{X(j)}} b_{X(j)\alpha} \\
& f_{j} \rightarrow e^{i\phi_{X(j)}} f_{X(j)}. \\
\end{split}
\end{equation}

The invariant gauge group (IGG) here is $Z_2$, hence $\phi_\textbf{e}=0$ or $\pi$ mod $2 \pi$. Considering all the algebraic constraints in Eq.\ref{multiplication}, the solutions of all the gauge transformations can be found as follows:

\begin{equation}
\label{psg}
\begin{split}
\phi_{T_1}	(x,y)&=0 \\
\phi_{T_2}(x,y)&=P_1 \pi x \\
\phi_{C_6}(x,y)&=\frac{1}{6} P' \pi +P_1\pi xy+\frac{P_1}{2}\pi y(y-1) \\
\phi_{\bar{\sigma}}(x,y)&=P_1\pi xy \\
\end{split}
\end{equation}
where $P_1=0,1$ and $P'=0,1,...,11$. Note that there are in total 24 solutions for PSG with $IGG=Z_2$. Since we are looking into $\pi$ flux states, $P_1=1$, and the PSG can be further simplified as:
\begin{equation}
\label{simplified_psg}
\begin{split}
\phi_{T_1}	(x,y)&=0 \\
\phi_{T_2}(x,y)&=\pi x \\
\phi_{C_6}(x,y)&=\frac{1}{3} P_3 \pi +\pi xy+\frac{1}{2}\pi y(y-1) \\
\phi_{\bar{\sigma}}(x,y)&=\pi xy \\
\end{split}
\end{equation}
where $P_3=0,1,2$. The number of solutions is reduced to 3 in triangular lattice.

\begin{figure}[t]
\centering
\includegraphics[width=3.4in]{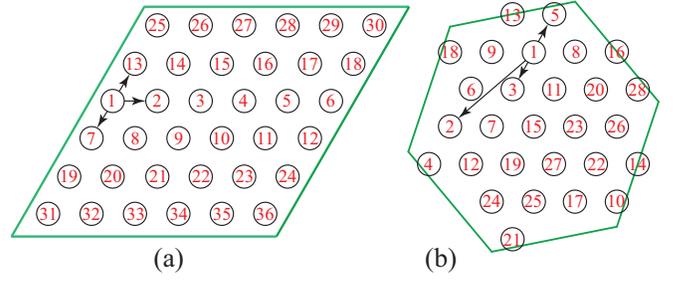}
\caption{ The representation of 2D samples into 1D chains with the site index increasing from 1 to N. (a) 36-sites sample. (b) 28-sites sample.\label{label}}
\end{figure} 

\begin{figure}[t]
\centering
\includegraphics[width=3.4in]{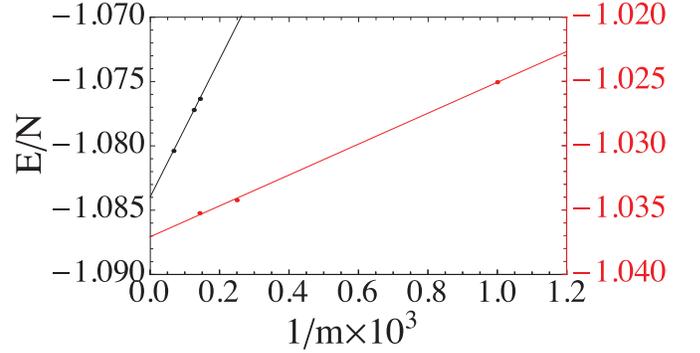}
\caption{ DMRG energy per site of t-J model with $J/t=0.6$ on 28-sites sample (Black) and 36 -sites sample (Red), as a function of limiting MPS matrix size $m$. \label{linear_fit}}
\end{figure} 

\section{\label{sec:level6} DMRG simulation details}
In our DMRG simulations, the size of the matrix product state (MPS) is limited due to computing core memory sizes. To obtain better convergence, the precise way of representing the 2D finite size samples with periodic boundary conditions matters. For 16-sites sample, since the size is sufficient small, the sites are labeled in the conventional way, i.e.  from top row to bottom row and from left to right in each row by site index $1,2,3,...,N$, where $N$ is the total number of sites in the sample. In this way of labeling, the maximal size, $m$, of MPS matrices saturates at $m=2274$. For 36-sites sample, since the number of sites is much bigger than 16-sites sample, we developed a better way to label the sites, as can be seen in Fig.\ref{label}(a), which can reduce the size of matrix product operator (MPO) matrix of the Hamiltonian from 282 to 138, and $m$ can be increased to 7000. For 28-sites sample, since the symmetry of the sample is quite different from rhombus shape samples, instead of labeling all the sites along the two Bravais lattice vectors $\vec{a}_{1(2)}$, we label them along $\vec{v}=\vec{a}_1+2\vec{a}_2$ and $\vec{a}_2$, as illustrated in Fig.\ref{label}(b), thus reducing the size of MPO from 218 to 98. To evaluate how well the DMRG simulations converge, we compare the energy obtained from different $m$ values as a function of $1/m$ and perform a linear fit extrapolation towards infinite $m$, see Fig.\ref{linear_fit}. The ground state energies obtained by the highest $m$ values approximate the extrapolated ground state energy quite well, with a reasonable offset of $<0.005$ compared to the extrapolated values at infinite $m$.

\end{document}